\documentclass[a4paper,11pt]{article}
\usepackage{amsmath, amssymb, amsfonts,indentfirst,graphicx,color,anysize,cite}
\usepackage{epigraph} 
\marginsize{3cm}{3cm}{2cm}{2cm}
\title{
	\includegraphics[width=0.35\textwidth]{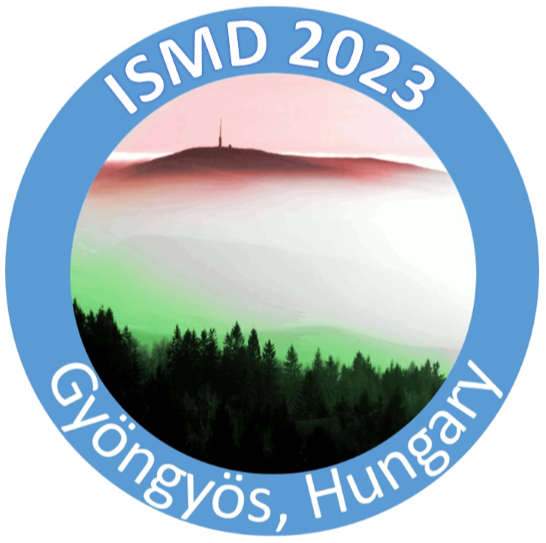}\\[1cm]
	\textbf{Model-independent Odderon results based on new TOTEM data on elastic  $pp$ collisions at 8 TeV}}
\author{{T.~Cs\"org\H{o}$^{1,2}$, T.~Nov\'ak$^{2,3}$, R.~Pasechnik$^4$, \underline{A.~Ster}$^1$, I.~Szanyi$^{1,2,5}$}\\[1ex]
	$^1$HUN-REN Wigner Research Center for Physics, \\
        Konkoly-Thege Mikl\'os \'ut 29-33, H-1121 Budapest, Hungary\\
	$^2$MATE Institute of Technology, K\'aroly R\'obert Campus, \\
        M\'atrai \'ut 36, H-3200 Gy\"ongy\"os, Hungary\\
     $^3$Department of Methodology for Business Analysis, Budapest Business     University, \\ Alkotm\'any \'ut 9-11, H-1054 Budapest, Hungary\\
    $^4$Division of Particle and Nuclear Physics, Department of Physics, \\
      Lund University, 221 00 Lund, Sweden\\
	$^5$Department of Atomic Physics, E\"otv\"os University, \\
        P\'azm\'any P. s. 1/A, H-1117 Budapest, Hungary\\
}
\begin{document}

\maketitle

\begin{abstract} 
%The statistically significant observation of the odderon exchange has been published in 2021 using the publicly available data on elastic $pp$ collisions at 7 TeV. 
%This analysis has been performed in a model-independent way based on the $H(x,s)$ scaling properties of the differential cross sections of elastic $pp$ scattering manifest in the TeV energy range. 
%Using  TOTEM data at $\sqrt{s} = 8$ TeV on the differential cross section and considering the $H(x,s)$ scaling properties of elastic $pp$ scattering.
Evaluating the $H(x,s|pp)$ scaling function of elastic proton-proton ($pp$) collisions from recent TOTEM data at $\sqrt{s} = 8$ TeV and comparing it with the same function of elastic proton-antiproton ($p\bar p$) data of the D0 collaboration 
at $\sqrt{s} = 1.96$ TeV, we find, from this comparison alone,  an  at least 3.79 $\sigma$  signal of  odderon exchange. 
If we combine this model-independently obtained result with that of a similar analysis but using TOTEM  elastic $pp$ scattering data at  $\sqrt{s} = 7$ TeV,
which resulted in an at least 6.26 $\sigma$  signal, 
the combined significance of odderon exchange increases to at least 7.08 $\sigma$.
%, model independently. 
Further combinations of various datasets in the TeV energy range are detailed in the manuscript. 
\end{abstract}

\setlength\epigraphwidth{.6\textwidth}
\epigraph {\small\textit{ ``No amount of experimentation can ever prove me right;\\ a single experiment can prove me wrong."}}{\textit{Albert Einstein}}

\section{Introduction}

In 1973, Lukaszuk and Nicolescu~\cite{Lukaszuk:1973nt} proposed that a noticeable crossing-odd contribution called odderon may be present in the %scattering 
amplitude of elastic proton-proton ($pp$) and proton-antiproton ($p\bar p$) scattering at asymptotically high energies. In the field theory of strong interactions, quantum chromodynamics (QCD),  odderon exchange corresponds to the $t$-channel exchange of a color-neutral gluonic compound state consisting of odd number of gluons~\cite{Bartels:1999yt}. For more than 20 years since the theoretical prediction of the odderon in the framework of QCD, and for more than 47 years since the odderon concept has been introduced in Regge phenomenology, the odderon remained elusive until the time of 2019-2021, due to lack of a definitive and statistically significant experimental evidence of odderon exchange.

A direct way to probe odderon exchange in elastic scattering is by comparing the differential cross-section of particle-particle and particle-antiparticle scattering at the same and sufficiently high energy~\cite{Jenkovszky:2011hu,Ster:2015esa}. The TeV energy scale is high enough to render the amplitudes for the  massive (Reggeon) exchange to negligibly small values. The differential cross-sections for elastic proton-proton as well as proton-antiproton scatterings are proportional to the modulus squares of the corresponding amplitudes. If the modulus squared amplitudes are different, then the two amplitudes cannot be equal. As the amplitude for odderon exchange is proportional to the difference between the amplitudes of the proton-antiproton and the proton-proton elastic scattering, 
it follows, that the amplitude for odderon exchange cannot vanish if the differential cross-sections of elastic proton-proton and proton-antiproton scatterings are significantly different at the TeV energy scale, as detailed in Ref.~\cite{Csorgo:2019ewn}.
%The differences are usually calculated by the $\chi^2$-method. In this special model-independent case we specify it in detail later in the manuscript.
A search performed at the ISR energy of $\sqrt{s}=53$ GeV in 1985~\cite{Breakstone:1985pe} resulted in an indication of the odderon at the 3.35 $\sigma$ significance level. That analysis, however, did not utilize all the available data in the overlapping acceptance of the $pp$ and $p\bar p$ measurements. Furthermore, at $\sqrt{s}=53$ GeV, Reggeon exchanges may play a significant role, rendering the odderon search at the ISR energies rather inconclusive.

As far as we know, the first anonymously peer-reviewed publication of a statistically significant, at least 6.26 $\sigma$ signal of odderon exchange was published in June 2020 by the authors of the present manuscript, in the proceedings of the ISMD 2019 (Santa Fe, NM, USA)~\cite{Csorgo:2020msw}. 
This refereed conference contribution was backed up in our February 2021 paper, with the same statistical significance of at least 6.26 $\sigma$ signal for odderon exchange~\cite{Csorgo:2019ewn}. Our Hungarian-Swedish team introduced a new scaling function that turned out to be energy independent in the LHC energies of $\sqrt{s}= 2.76 - 7.0$ TeV in elastic proton-proton ($pp$) collisions, based on a model-independent, direct data to data comparison~\cite{Csorgo:2019ewn}. 
Our idea was based on the existence of a diffraction cone at small values of the squared four-momentum transfer, the Mandelstam variable $t$, characterized by an exponential fall. By plotting the $pp$ and $p\bar p$ data as a function of a dimensionless variables $x = - B(s) t $, where $B$ is the characteristic scale of this exponential fall, and by dividing the differential cross-section by $B \, \sigma_{\rm el}$, where $\sigma_{\rm el}$ is the integral of the differential cross-section, (dominated by the contribution of the diffractive cone), a dimensionless $H(x,s)$ scaling function is obtained. In the region of the diffractive cone, this
scaling function is approximately a structureless exponential, $H(x) \approx \exp(-x)$. As the diffractive cone is located at small values of $t$, this $H(x)$ scaling removes the trivial $s$-dependences of the differential cross-section, where the Mandelstam variable $s$ is the squared center of mass energy.
Thus such a {\it $H(x)$-scaling} removes the trivial energy dependent terms, due to the known $s$-dependence of the elastic slope $B(s)$, the elastic and total cross-sections $\sigma_{\rm el}(s)$ and $\sigma_{\rm tot}(s)$, and the real-to-imaginary ratio $\rho(s)$~\cite{Csorgo:2019fbf}. 
It follows, that the amplitude for odderon exchange cannot vanish if the dimensionless  
$H(x)$ scaling functions of elastic proton-proton and proton-antiproton scatterings are significantly different at the TeV energy scale ~\cite{Csorgo:2019ewn}.
%In this article we have proven that if we renormalize the differential cross-sections by their simultaneously published diffrative slope parameters (B) and the total $\sigma_{el}$'s, i.e., $H(x) = 1/(B \sigma_{el}) \,\, d\sigma/dt$ then we eliminate energy dependence, at least in a few factors of collision energy.

%When comparing this so-called energy independent $H(x | pp)$ scaling function with the energy dependent $H(x,s | p\bar p)$ scaling function of elastic proton-antiproton $(p \bar p)$ collisions, evaluated by our group  directly from public domain, published  differential cross-section data of elastic $pp$ and $p\bar p$ scattering data model-independently.

These results as well as the model-{\it dependent} investigation of the domain of validity of the $H(x,s|pp)$ scaling
have been seconded by a theoretical paper of T. Cs\"org\H o, and I. Szanyi~\cite{Szanyi:2022ezh}, increasing the statistical significance of the observation 
of odderon exchange to at least 7.08 $\sigma$. 
At the same time, this  model-dependent investigation found that the domain of validity in $s$ of the $H(x,s|pp) = H(x,s_0|pp)$ scaling (where $s_0$ stands for a fixed energy scale in the TeV region, e.g. $\sqrt{s_0} = 7 $ TeV)  extends also to the top Tevatron energies of $\sqrt{s} = 1.96$ TeV~\cite{Szanyi:2022ezh}. This theoretical work utilized a validated model, proposed in its first form by A. Bialas and A. Bzdak~\cite{Bialas:2006qf},
however, the original model lacked a real part hence the possibility of odderon exchange. However, the so-called Real-extended Bialas-Bzdak (ReBB) model of Ref.~\cite{Nemes:2015iia} 
fixed these shortcomings and has been utilized in Ref.~\cite{Szanyi:2022ezh} to extrapolate not only the elastic proton–proton scattering data from the LHC energies of $\sqrt{s} = 2.76$ and 7 TeV to the D0 energy of $\sqrt{s} =  1.96$ TeV but also to extrapolate the elastic proton–antiproton scattering data from $\sqrt{s} = 0.546$ and 1.96 TeV of the UA4 ~\cite{UA4:1983mlb,UA4:1985oqn} and the D0 Collaborations~\cite{Abazov:2012qb}
to the LHC energies of 2.76 TeV and 7 TeV. Evaluating the proton–proton data with a model increased the uncertainty and decreased the odderon signal from proton–proton scattering data alone, but this decrease was well over-compensated with the ability of the model to evaluate theoretically the proton–antiproton scattering at all the LHC energies. Overall, this procedure resulted in a model-dependent  increase of the statistical significance from 
odderon exchange from 6.26 to 7.08 $\sigma$~\cite{Szanyi:2022ezh} as published in final form in July 2021, but limited the comparison to the diffractive minimum and maximum region in the four-momentum transfer range, to the 
verified domain of  validity
of the ReBB model. More recently, these results were extended to 
the new TOTEM data on elastic $pp$ scattering at $\sqrt{s} = 8$ TeV, published in March 2022~\cite{TOTEM:2021imi} in Ref.~\cite{Szanyi:2022ezh}. 
When TOTEM data on elastic $pp$ collisions at $\sqrt{s} = 8$, 7, and 2.76 TeV are analyzed simultaneously with D0 data at $\sqrt{s} = $ 1.96 TeV in the framework of the ReBB model, a combined statistical significance greater than 32.4 $\sigma$ can be achieved, rendering the statistical significance of odderon observation, in any practical terms, to a certainty~\cite{Szanyi:2022ezh}.

A series of important papers has been published by the TOTEM Collaboration investigating the properties of elastic $pp$ scattering in the LHC energy range between $\sqrt{s} = 2.76$ and $13$ TeV in Refs.~\cite{Antchev:2017dia,Antchev:2017yns,Antchev:2018edk,Antchev:2018rec}. Most recently, the latest measurement performed by TOTEM at $\sqrt{s} =  8$ TeV \cite{TOTEM:2021imi} extended the earlier analysis up to $|t|$-values of 1.9 GeV$^2$ \cite{Antchev:2015zza}. 
An increase of the total cross section, $\sigma_{\rm tot}(s)$, associated with a decrease of the real-to-imaginary ratio, $\rho(s)$, with energy, first identified at $\sqrt{s} = 13$ TeV \cite{Antchev:2017dia,Antchev:2017yns} indicated a possible odderon effect triggering an intense discussion in the literature (see 
e.g.~Refs.~\cite{Khoze:2017swe,Samokhin:2017kde,Csorgo:2018uyp,Broilo:2018qqs,Pancheri:2018yhd,Goncalves:2018nsp,Selyugin:2018uob,Khoze:2018bus,Broilo:2018els,Troshin:2018ihb,Dremin:2018uwt,Martynov:2018nyb,Martynov:2018sga,Shabelski:2018jfq,Khoze:2018kna,Hagiwara:2020mqb,Contreras:2020lrh,Gotsman:2020mkd}). 
The persistent diffractive minimum-maximum structure in the $t$-dependent profile of $d\sigma/dt$ in elastic $pp$ collisions observed by TOTEM at $\sqrt{s}$ = 2.76, 7, 8 and 13 TeV, and the lack of such structure in elastic $p\bar p$ collisions measured by D0~\cite{Abazov:2012qb} at $\sqrt{s} = 1.96$ TeV indicate a qualitatively clear odderon effect. The possibility of utilizing experimental data at the TeV energy scale for a search for a statistically significant odderon exchange 
has been proposed by Jenkovszky et al. in Refs.~\cite{Ster:2015esa,Csorgo:2018uyp}.
In 2020, the TOTEM collaboration made a qualitative conclusion about odderon exchange in Ref.~\cite{Antchev:2018rec} as follows: {\it ``Under the condition that the effects due to the energy difference between TOTEM and D0 can be neglected, the result provides evidence for a colourless 3-gluon bound state exchange in the $t$-channel of the $pp$ elastic scattering''}. However, no statistical significance for this observation has been evaluated in Ref.~\cite{Antchev:2018rec}.

More recently, in August 2021, a properly quantified statistical significance of the odderon signal has been published by the TOTEM and D0 Collaborations \cite{D0:2020tig} employing different methods and techniques, obtaining an at least 5.2 $\sigma$ combined statistical significance for an almost model-independent observation of odderon exchange, a first statistically significant result obtained by two experimental collaborations. This result was based on the extrapolation of 
TOTEM experimental data of the differential cross-section of elastic $pp$ scattering from $\sqrt{s} = $ 13, 8, 7 and 2.76 TeV down to 1.96 TeV using an almost model-independent technique and comparing the results with D0 data in a limited four-momentum transfer range, resulting in an at least 3.4 $\sigma$ signal for odderon exchange. TOTEM also has measured the pair of the total cross-section and the real-to-imaginary ratio $(\sigma_{tot},\rho)$ and compared it with a set of models without odderon exchange. When a partial combination of the TOTEM $(\sigma_{tot},\rho)$ measurements is done at $\sqrt{s} = 13$ TeV, the obtained partial significances range between 3.4 and 4.6 $\sigma$ for the considered models. The full combination of the signal of odderon exchange from TOTEM $(\sigma_{tot},\rho)$ measurements at $\sqrt{s} = 13$ TeV with the signal of the comparision of extrapolated TOTEM $pp$ data to $\sqrt{s} = 1.96$ TeV with a subset of 8 out of 17 D0 datapoints~\cite{Abazov:2012qb}
on elastic $p\bar p$ scattering leads to total significances ranging from 5.2 to 5.7 $\sigma$ for $t$-channel odderon exchange for each of the considered models~\cite{D0:2020tig}.

The validity of this D0-TOTEM proof of odderon exchange has been questioned in several published papers by now. Most importantly,  the ATLAS collaboration~\cite{ATLAS:2022mgx} published a significantly different total cross-section hence a significantly different pair of $(\sigma_{tot}, \rho)$
at $\sqrt{s} = 13 $ TeV, questioning the significance of the signal of odderon exchange from these low $-t$ observations. 
Such a significant incompatibility between the ATLAS and TOTEM measurements of total cross-section and the ratio of real to imaginary part of the scattering amplitude, that is between the pairs of $(\sigma_{tot},\rho)$  at $\sqrt{s} = 13$ TeV as published by the ATLAS and by the TOTEM experiments has recently been confirmed by recent results of Petrov and Tkachenko in Refs.~\cite{Petrov:2022fsu,Petrov:2023mww,Petrov:2023lho}.
Furthermore, Donnachie and Landshoff~\cite{Donnachie:2019ciz,Donnachie:2022aiq} stressed the point that phase of an elastic scattering amplitude is related to its energy variation, and as a consequence, they have questioned the validity of the D0-TOTEM signal of odderon exchange at $t= 0$. 
Petrov and Tkachenko obtained  results similar to that of Donnachie and Landshoff in Ref.~\cite{Petrov:2022fsu}, 
suggesting that the systematic error on the determination of the $\rho$ parameter at $\sqrt{s} = 13$ TeV might have been slightly but significantly underestimated by TOTEM in Ref.~\cite{Antchev:2017yns}. Let us mention that Refs.~\cite{Csorgo:2020msw,Csorgo:2019ewn} scale out the $t = 0$ observables from their analysis of the
$H(x,s|pp)$ scaling functions,
while the low-$t$ domain has been explicitely excluded from finding a statistically significant signal of odderon-exchange in Refs.~\cite{Csorgo:2021epjc,Szanyi:2022ezh}.
Hence these odderon discovery papers are not affected by the above mentioned criticisms of the D0-TOTEM publication of a statistically significant, at least 5.2 $\sigma$ experimental observation of odderon exchange.

Let us note, that the  ATLAS data that were measured  in the diffraction cone at low $-t$ at $\sqrt{s} = $ 7, 8 and 13 TeV in Refs.~\cite{ATLAS:2014vxr,ATLAS:2016ikn,ATLAS:2022mgx}, respectively,
collapse to the same $H(x)$ scaling function, that is close to a structureless $H(x) \approx \exp(-x)$. 
Hence, the method discussed in this manuscript cannot be utilized for a search of odderon effects
in these ATLAS datasets.

In addition to the criticism of the D0-TOTEM method of using $t=0$ data at $\sqrt{s} = 13$ TeV for the observation of $t$-channel odderon exchange, Cui and collaborators~\cite{Cui:2022dcm} utilized a mathematical approach based on interpolation via continued fractions enhanced by statistical sampling and suggested that a model-independent extrapolation of TOTEM experimental data of the differential cross-section of elastic $pp$ scattering from $\sqrt{s} = $ 13, 8, 7 and 2.76 TeV down to 1.96 TeV and comparing the results with D0 data in a limited four-momentum transfer range, results in only an at least 2.2 $\sigma$ signal for odderon exchange. 
This result alone decreases the significance of the D0-TOTEM combined result for odderon exchange from an at least 5.2 $\sigma$ to an at least 4.0 $\sigma$ signal for odderon exchange~\cite{Cui:2022dcm}, suggesting that the D0-TOTEM method of proving the significance of odderon exchange is only on the level of  an indication (defined as a significance between 3.0 $\sigma $ and  5.0 $\sigma$), but falls a little bit too short from being experimentally conclusive, definitive discovery as the corrected value falls short of the discovery threshold of 5 $\sigma$. Such an at least 2.2 $\sigma$ signal for odderon exchange from extrapolating the
TOTEM measured differential cross-sections of $\sqrt{s} = $ 8, 7 and 2.76 TeV down to 1.96 TeV confirms the model-dependent results of Refs.~\cite{Csorgo:2021epjc,Szanyi:2022ezh} as well. 

A response to these published criticisms was given by the talk of K. Österberg, the physics coordinator of the TOTEM experiment at the ISMD 2023 conference in Gyöngyös, Hungary extending an earlier D0-TOTEM response given also by him at the 2021  Low-x Workshop at La Biodola, Italy ~\cite{Osterberg:2022qoy,Osterberg:2023Gyon}.   We have good reasons to expect that a detailed D0-TOTEM response to the above criticisms  will be submitted for a publication as soon as reasonably possible.  Furthermore, we also second the suggestion of Petrov and Tkachenko in Ref.~\cite{Petrov:2023lho}, calling for a joint ATLAS-TOTEM analysis to sort out the differences between their low-$t$ measurements at $\sqrt{s} =13$  TeV, proposing also a comparison of ATLAS and TOTEM data at low $-t$ at $\sqrt{s} = 7$ TeV~\cite{TOTEM:2013lle,Antchev:2013iaa,ATLAS:2014vxr} and the same comparison also at $8$ TeV~\cite{TOTEM:2016lxj,ATLAS:2016ikn}.

Let us note, that this  ongoing debate in the literature focuses on questioning the validity of certain D0-TOTEM proofs of a statistically significant observation of odderon exchange, but this debate does not question the existence and statistical significance of odderon exchange in all the four published papers on this topic at this energy scale.  The statistically significant, well above the  5.0 $\sigma$ observations of a $t$-channel odderon exchange, as published in Refs.~\cite{Csorgo:2020msw,Csorgo:2019ewn} as well as in Refs.~\cite{Csorgo:2021epjc,Szanyi:2022ezh}, have  not been affected by the above criticism and have not been challenged so far in other publications either, as far as we know. Furthermore, the statistical significance of odderon exchange as determined from the ReBB model analysis has been increased by taking into account the new 8 TeV datapoints of the TOTEM experiment by I. Szanyi and T. Csörgő:  In any practical terms, within the framework of the ReBB model, the signal for odderon exchange in the limited $0.37 \leq -t \leq 1.2$ GeV$^2$
 and $1.96 \leq \sqrt{s} \leq 8$ TeV kinematic region is so large that it amounts to not a probability, but a certainty~\cite{Szanyi:2022ezh}.

%In the mean time the statistical significance of odderon exchange from the ReBB model analysis has been increased by taking into accont the new 8 TeV datapoints of the TOTEM experiment by I. Szanyi and T. Csöre{Csorgo:2021epjc}. So at least 3 papers are published at the time of writing this manuscript that prove the based on published experimental data including that of  D0 and TOTEM a statistically  significant contribution of odderon exchange in the TeV energy range. 

In the present manuscript, we summarize our {\it model-independent} analysis of the statistical significance of the odderon observation using the recently published~\cite{TOTEM:2021imi} and extended~\cite{D0:2020tig} $\sqrt{s} = 8 $ TeV data set  of the TOTEM Collaboration in elastic $pp$ collisions, together with earlier data from D0~\cite{Abazov:2012qb} and TOTEM~\cite{TOTEM:2013lle,Antchev:2018edk,Antchev:2018rec} Collaborations, extending our earlier scaling studies of the differential $pp$ scattering cross section at TeV energies up to 8 TeV. 

Our approach is model-independent in the sense that it does not rely on any fitting function or any theoretical input, it uses only linear and log-linear interpolation techniques between neighbouring datapoints, to allow for data to data comparison at the same values of the horizontal axes (using the scaling variable $x = - t B$). As we compare pairwise the $H(x,s|pp)$ scaling functions constructed at different energies based only upon the available data and look for statistically significant differences within any pair of TeV-scale $pp$ and $p\bar p$ data sets depending on the collision energy, we need to utilize rebinning that also includes model-independent linear interpolation methods, as detailed in Ref.~\cite{Csorgo:2019ewn}, where the basic concepts and methodology have been explained in detail.

Note that we have determined the domain of validity of the applicability of the $H(x, s |pp)$ scaling at $\sqrt{s} = 1.96$ TeV so far model-dependently only,
in Ref.~\cite{Csorgo:2019ewn}, based on the ReBB model of Refs.~\cite{Nemes:2015iia,Csorgo:2021epjc,Szanyi:2022ezh}. The domain of validity of this $H(x|pp)$ scaling has been found to include $\sqrt{s} = 1.96$,  $2.76$, $ 7$ and $ 8$ TeV in a model-independent way as well, as presented at various conferences, e.g.~\cite{Csorgo:2023Prot}, but these results go way beyond the scope of the present manuscript and will be 
detailed elsewhere.

%Stop here for now (T. Cs.)

\section{$H(x)$ scaling of 2.76, 7 and 8 TeV $pp$ data of TOTEM}

Our analysis is based on our recent discovery of a novel scaling law of elastic $pp$ scattering at TeV energies, referred to as $H(x)$ scaling, as described in ref~\cite{Csorgo:2019ewn}. The scaling function $H(x,s|pp)$ is defined for elastic proton-proton ($pp$) collisions as follows:
\begin{eqnarray}
 H(x,s | pp ) &  =  & \frac{1}{B \, \sigma^{pp}_{\rm el}} \frac{d\sigma^{pp}}{dt} , \label{scaling}  \\
  x  & = &  -t \, B, \\
 \sigma^{pp}_{\rm el} & = &  \int_{0}^\infty d|t| \,\, \frac{d\sigma^{pp}}{dt}, \\
  B  & = & \frac{d}{dt} \ln \frac{d\sigma^{pp}}{dt}\Big|_{t\to 0}
\end{eqnarray}
A similar function $H(x,s|p\bar p)$ can be introduced for elastic proton-antiproton ($p \bar p)$ collisions. In Ref.~\cite{Csorgo:2019ewn} we have shown,
that $H(x,s | pp ) = H(x | pp )$ is independent of the energy  in elastic $pp$ collisions, utilizing direct data to data comparison of elastic $pp$ data
of the TOTEM Collaboration at $\sqrt{s} = 2.76$ and $7$ TeV, as described in Refs.~\cite{Antchev:2018rec,TOTEM:2011vxg,TOTEM:2013lle}. In Ref.~\cite{Csorgo:2019ewn} we have also shown, in a model-dependent way, that this scaling extends down to $\sqrt{s} = 1.96$ TeV, and we have also found that the scaling violations are significant
already at $\sqrt{s} = 13$ TeV. In the present study we investigate if the energy independence of this scaling is, within experimental errors, valid for the recently published TOTEM data at $\sqrt{s} = 8$ TeV, or not, within the experimental errors as
determined by the TOTEM collaboration in  Ref. ~\cite{TOTEM:2021imi}. This question is particularly interesting as 
the elastic to total cross-section ratio increases with increasing
energies, and at the LHC energies between 7  and 13 TeV it crosses
significantly the important limit of  $(1 + \rho^2) \sigma_{\rm el}/\sigma_{\rm tot} = 1/4$, see Fig. 4 of Ref.~\cite{Csorgo:2019fbf}. 
This ratio reaches such a critical value in the region of $\sqrt{s} = 2.76 $ -- $7.0 $ TeV and it clearly exceeds it at $\sqrt{s} = 13.0$ TeV, where we find statistically significant violations of the $H(x,s|pp) = H(x,s_0|pp) $ scaling law. The search for the onset of the violation of $H(x,s|pp) = H(x,s_0|pp) = H(x,pp)$ scaling motivates our current investigation.

The $H(x | pp)$ scaling function was found to increase the statistical significance of the odderon signal, based on a direct comparision of $H(x,s | pp ) = H(x | pp )$
and $H(x,s | p \bar p )$ at $\sqrt{s} = 1.96$ TeV,
due to at least two reasons. First of all, the overall normalization uncertainty, the largest source of the systematic error in differential cross-section measurements, cancels from $H(x,s|pp)$ (as well as from $H(x,s|p\bar p)$ ) . Secondly,  this $H(x | pp)$ scaling law reduces the collision energy related uncertainties in the data which consequently increases the precision of the extracted quantities. We may also mention that in the diffractive cone, expected to be valid at least up to $x = - B t \ll 1$, the $H(x) $ scaling function is expected to start as $H(x) \approx \exp(-x)$, with a well-defined normalization at the optical point: 
\begin{equation}
    H(x = 0,s |pp ) = H(x = 0, s | p\bar p )  = 1,
\end{equation}
as follows from the definitions of this scaling function.
This property of the scaling functions removes the uncertainty related to possible small differences between the optical points of the differential cross-sections in elastic $pp$ and $p\bar p$ collisions that may exist between $pp$ and $p\bar p$ elastic collisions even if they are measured at the same center of mass energies. 
%%%%%%%%%%%%%%%%%%%%%%%%%%%%%%%%%%
\begin{figure*}[hbt]
\begin{center}
\begin{minipage}{1.0\textwidth}
 \centerline{
 \includegraphics[width=0.48\textwidth]{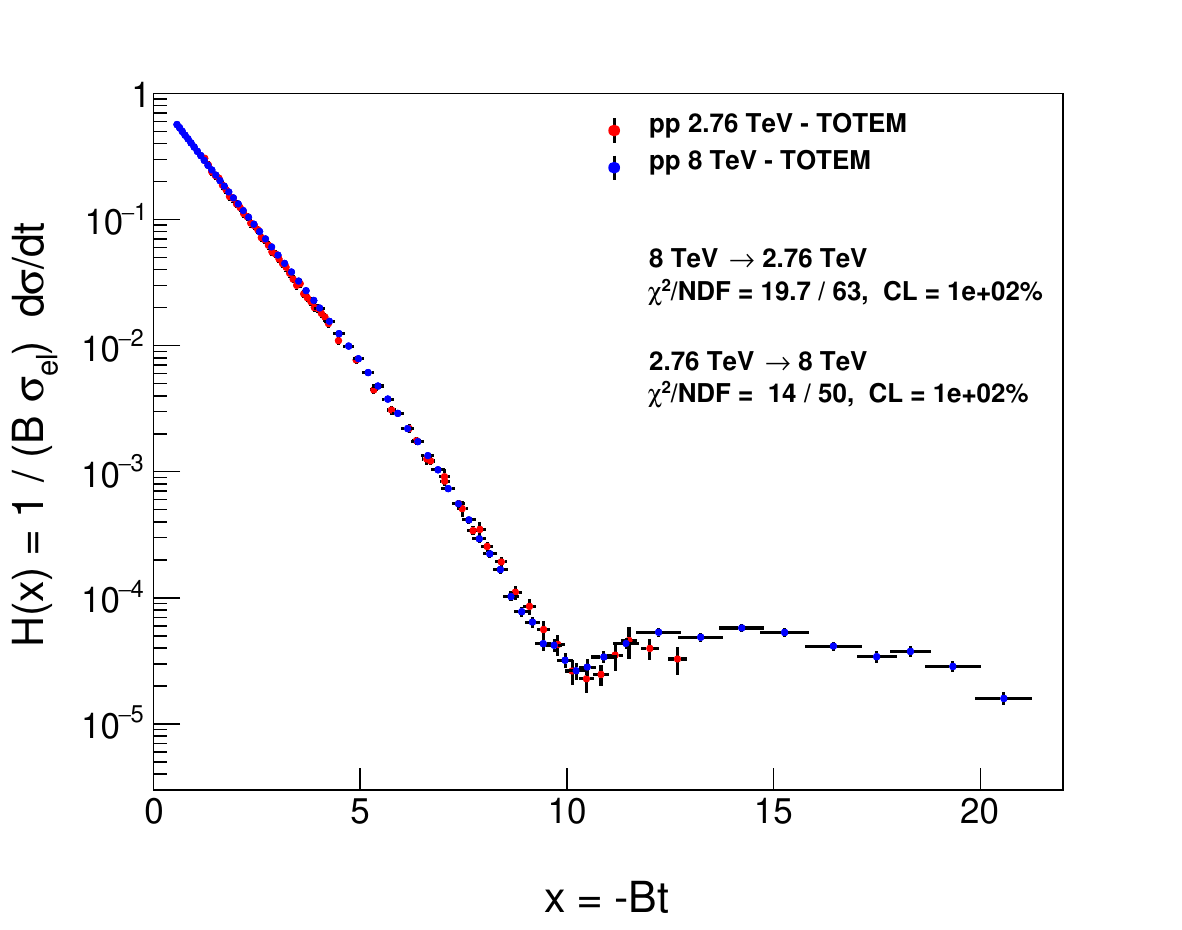}
 \includegraphics[width=0.48\textwidth]{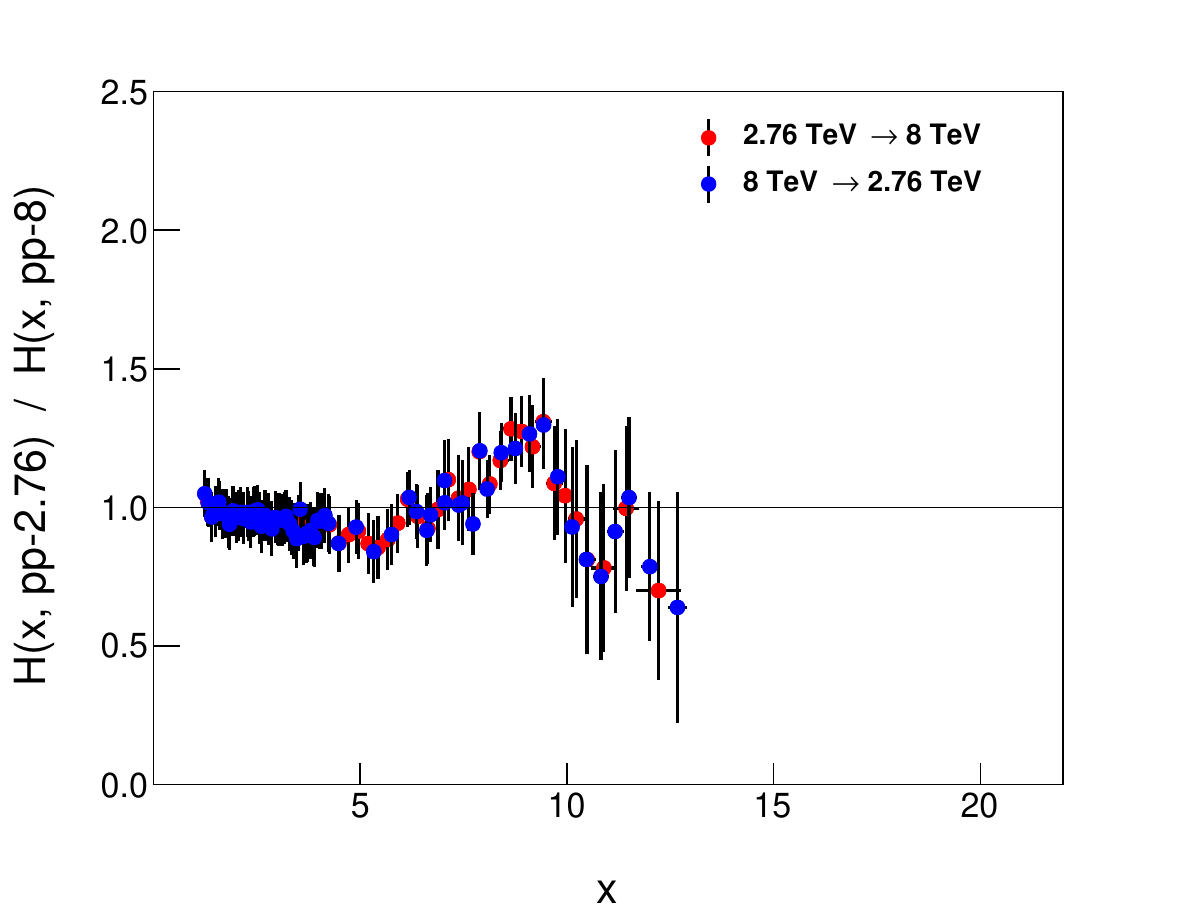}
 }
\vspace{-0.5truecm}
\end{minipage}    
\end{center}
\caption{ 
{\it Left panel} indicates that for $pp$ elastic scattering the $H(x)$ scaling function for $x = - t B$ is energy independent in the energy range of $\sqrt{s} = 2.76 - 8$ TeV. 
The notation is the same as in Ref.~\cite{Csorgo:2019ewn} for the data points and the decomposition of the errors to point-to-point fluctuating, type A errors (vertical and horizontal lines),
to point-to-point correlated, type B errors (grey bars), noting that the overall normalization (type C) errors cancel from the $H(x,s)$ scaling function. The $X\to Y$ notations are indicating the direction of the projection, as detailed in  Ref.~\cite{Csorgo:2019ewn}.
%Vertical and horizontal lines on each point stand for the corresponding type A errors. Grey vertical bars represent the type B (vertical and horizontal) errors. 
%
{\it Right panel} indicates that for $pp$ elastic scattering the ratio of the scaling functions $H(x,s_1)/H(x,s_2)$, where $x = - t B$, $\sqrt{s_1}=2.76$ TeV and $\sqrt{s_2}=8$ TeV, is {\it not inconsistent} with unity within statistical errors, due to the energy independence of the $H(x,s)$ scaling in the $2.76 \le \sqrt{s_{1,2}} \le 8$ TeV energy range. Here, $X\to Y$ denotes direction of the projections by exponential interpolation between two adjacent data points of the data set $X$ to get its $H(x)$ at the same $x$ where the other data set $Y$ was measured, so that we can compare them via $\chi^2$-method as detailed in the text
and in Ref.~\cite{Csorgo:2019ewn}.
}
\label{fig:H(x)-Odderon-1}
\end{figure*}
%%%%%%%%%%%%%%%%%%%%%%%%%%%%%%%%%%
\begin{figure*}[hbt]
\begin{center}
\begin{minipage}{1.0\textwidth}
 \centerline{
 \includegraphics[width=0.48\textwidth]{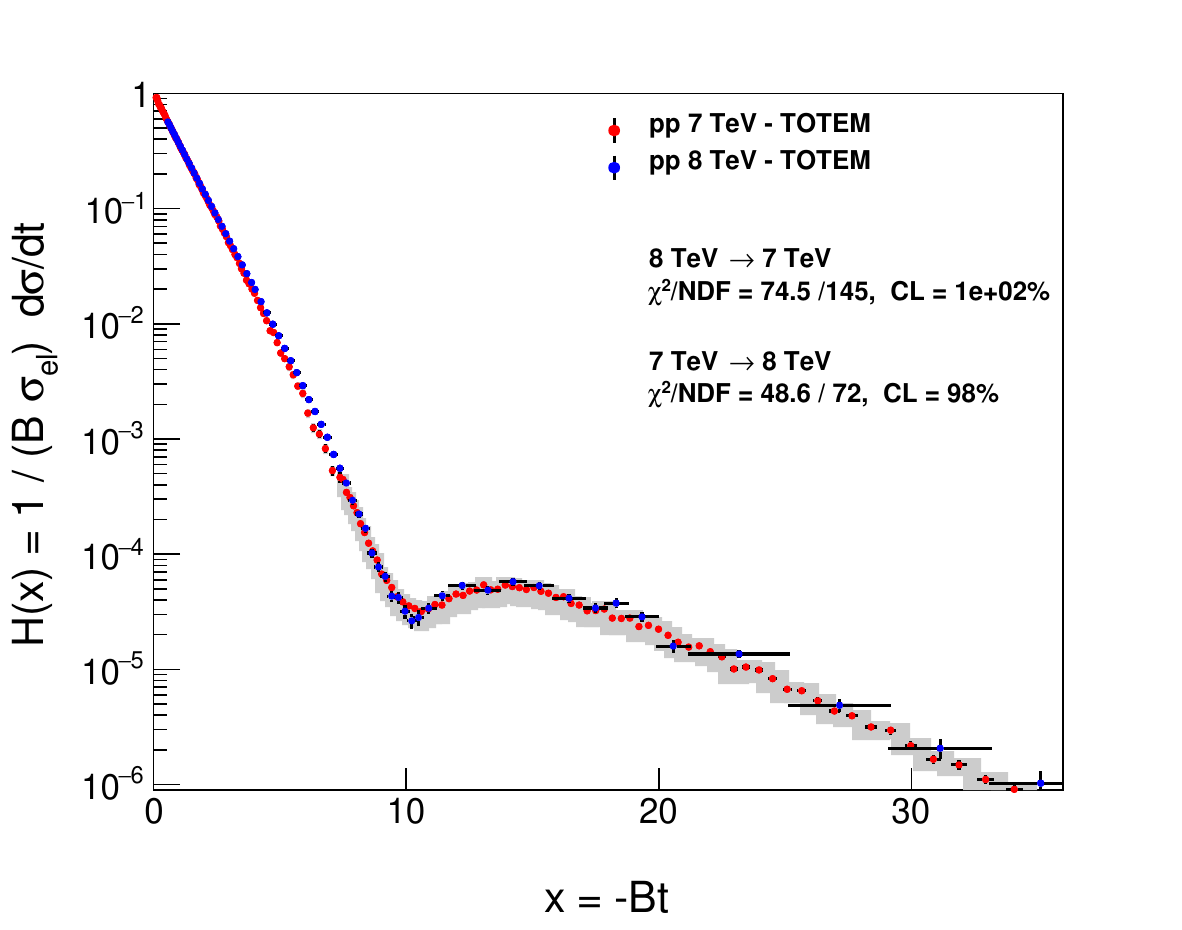}
 \includegraphics[width=0.48\textwidth]{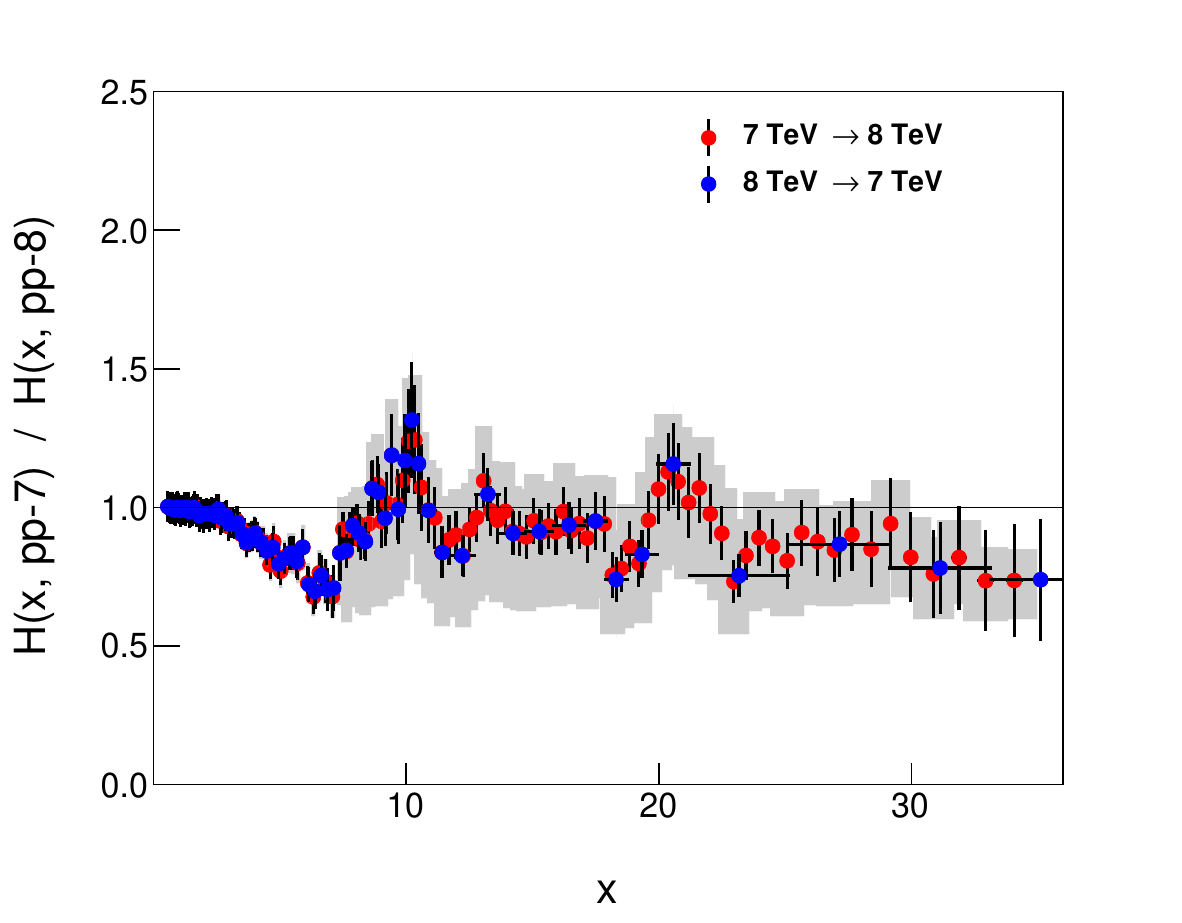}
 }
\vspace{-0.5truecm}
\end{minipage}    
\end{center}
\caption{ 
Left panel indicates that for $pp$ elastic scattering the $H(x)$ scaling function for $x = - t B$ is energy independent in the energy range of $\sqrt{s} = 7 - 8$ TeV. %Vertical and horizontal lines on each point stand for the corresponding type A errors. Grey vertical bars represent the type B (vertical and horizontal) errors. 
The notation is the same as in Fig. \ref{fig:H(x)-Odderon-1} as detailed in 
%for the datapoints and the decomposition of the errors to point-to-point fluctuating, type A errors (vertical and horizontal lines),
%to point-to-point correlated, type B errors (grey bars) is the same as in Ref.~\cite{Csorgo:2019ewn}. The $X\to Y$ notations are indicating the direction of the projection, as detailed in  
Ref.~\cite{Csorgo:2019ewn}.
Right panel indicates that for $pp$ elastic scattering the ratio of the scaling functions $H(x,s_1)/H(x,s_2)$, where $x = - t B$, $\sqrt{s_1}=7$ TeV and $\sqrt{s_2}=8$ TeV, is {\it not inconsistent} with unity within statistical errors, due to the energy independence of the $H(x,s)$ scaling function in the $2.76 \le \sqrt{s_{1,2}} \le 8$ TeV energy range. 
}
\label{fig:H(x)-Odderon-2}
\end{figure*}
%%%%%%%%%%%%%%%%%%%%%%%%%%%%%%%%%%
%\begin{figure*}[hbt]
%\begin{center}
%\begin{minipage}{1.0\textwidth}
% \centerline{
% \includegraphics[width=0.48\textwidth]{figs/Fig-compare.png}
% \includegraphics[width=0.48\textwidth]{figs/Fig-compare_savos.png}
% }
%\vspace{-0.5truecm}
%\end{minipage}    
%\end{center}
%\caption{ 
%\color{red}
%This panel indicates a partial disagreement beyond the published experimental errors in the low range of $B t$. One can see that a new type of systematic errors appears at around $x=7.5$.
%}
%\label{fig:comparison}
%\end{figure*}
%%%%%%%%%%%%%%%%%%%%%%%%%%%%%%%%%%
\begin{figure*}[hbt]
\begin{center}
\begin{minipage}{1.0\textwidth}
 \centerline{
 \includegraphics[width=0.48\textwidth]{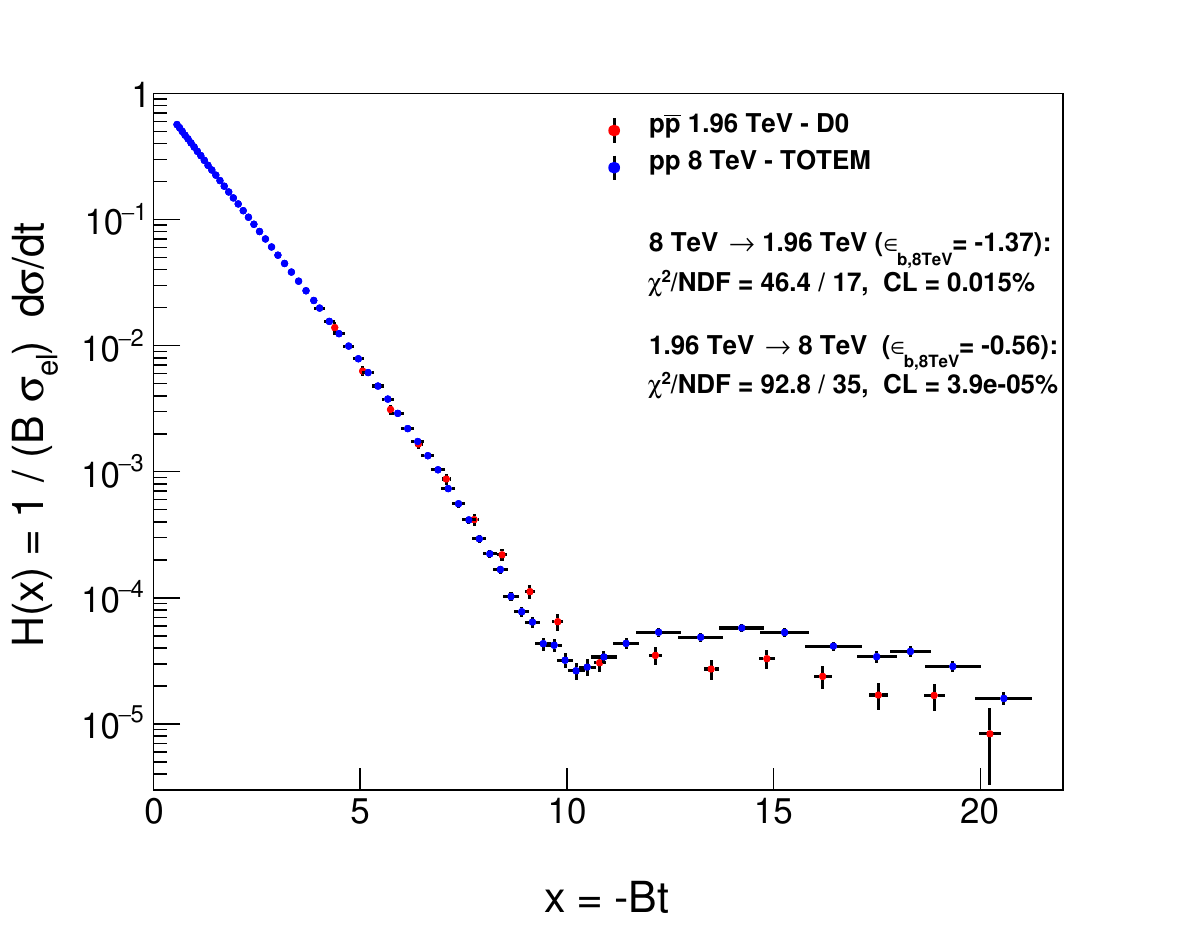}
 \includegraphics[width=0.48\textwidth]{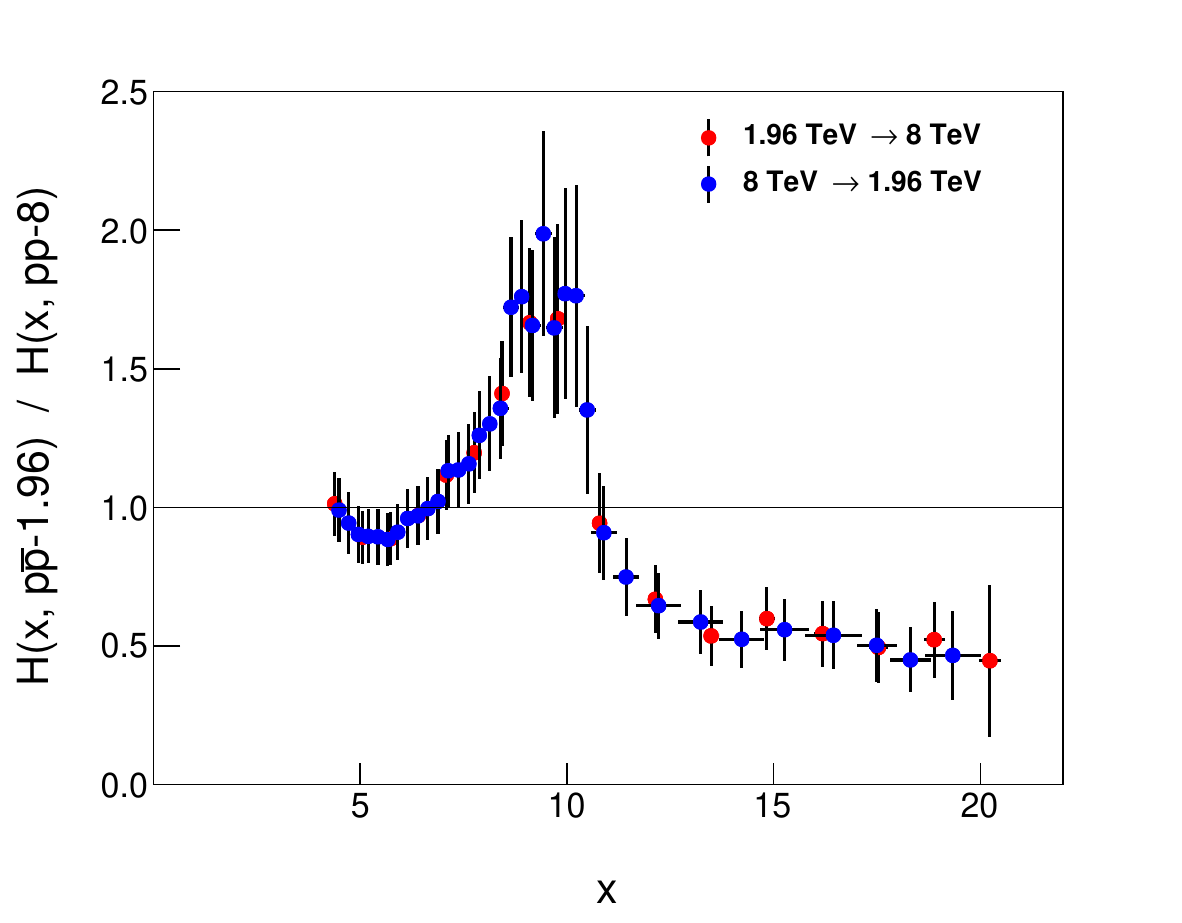}
 }
\vspace{-0.5truecm}
\end{minipage}    
\end{center}
\caption{ 
Left and right panels indicates a statistically significant difference between the $H(x)$ scaling functions for elastic $pp$ collisions at $\sqrt{s} = 8$ TeV and that of $p\bar p$ collisions at $\sqrt{s} = 1.96$ TeV at the level of at least 3.79 $\sigma$ and 5.1 $\sigma$, depending on the direction of projection, respectively. 
%However, both values get larger (above 5$\sigma$) if better estimated horizontal errors are applied on the $8$ TeV data set as discussed in the text. 
Notations are the same as in Fig.~\ref{fig:H(x)-Odderon-1}, and detailed in Ref.~\cite{Csorgo:2019ewn}.
}
\label{fig:H(x)-Odderon-3}
\end{figure*}
%%%%%%%%%%%%%%%%%%%%%%%%%%%%%%%%%%

%The scaling function has been found to be energy independent, i.e. $H(x,s_1) = H(x,s_2)\equiv H(x)$, in elastic $pp$ collisions, at least, in the energy domain of a few TeV (more specifically, at $\sqrt{s} = 2.76 - 8$ TeV) within the acceptance of TOTEM measurements at $2.76$, $7$ and $8$ TeV. 
%This is explicitly demonstrated 
In the left panels of Fig.~\ref{fig:H(x)-Odderon-1} and Fig.~\ref{fig:H(x)-Odderon-2}, we demonstrate the energy independence of the { $H(x | pp)$-scaling}
within statistical errors, where the TOTEM datasets at $\sqrt{s} = 2.76$ and $8$ TeV as well as at $\sqrt{s} = 7$ and $8$ TeV are compared pairwise,
respectively. The agreement between these datasets corresponds to a confidence level (CL) of at least 98\% . 

%A quantitative Odderon effect is then determined by a projection of the $pp$ and $p\bar p$ data and the corresponding uncertainties to the same $x(s)$ values found at different energies.

However, one should note that in a local low $x$ (low $|t|$) interval one can see larger deviations than the reported data errors suggest:
In case of the comparison of the $H(x,s|pp)$ scaling functions at $ \sqrt{s} = 7$ TeV and $8$ TeV, we observe a partial disagreement beyond the published experimental errors in a small $x$ range of about $x = [3.48 - 7.45]$ on Fig.~\ref{fig:H(x)-Odderon-2} corresponding to $|t| = [0.175 - 0.375]$ GeV$^2$, which points toward a possible adjustment problem of the two parts of the data sets measured with different optics ($\beta^* = 3.5 $ m and $\beta^* = 90 $ m and published separately in Tables 5 and 4 of Ref.~\cite{TOTEM:2013lle}, respectively)
and suggests the need of a further, more detailed investigation of the published systematic errors in the low $|t| = [0.175 - 0.375]$ GeV$^2$ range of the dataset measured by TOTEM at $\sqrt{s} = 7$ TeV with the $\beta^* = 90 $ m LHC optics~\cite{TOTEM:2013lle}.

%of the lower and higher arm detectors and 

%This is the first direct observation of systematics in the $7$ TeV low $-t$ dataset that are beyond the reported errors. 
Difficulties of  describing this part of the TOTEM  data at $\sqrt{s} = 7$ TeV have been reported in several earlier analyses, 
but these problems were typically attributed to the insufficiency of the applied analysis methods, for examples see Figs. 1 and 2 of Ref.~\cite{Nemes:2015iia} or Fig. 2 of Ref. ~\cite{Ster:2015esa}.
So, we suggest caution when using of the $\sqrt{s} = 7$ TeV TOTEM  data in the $|t| = [0.175 - 0.375]$ GeV$^2$ four-momentum transfer region. 

Fortunately, this $-t$ range, when mapped to $x = - Bt$, overlaps only marginally with the D0 acceptance and data, at three points only, 
not impacting our conclusions about a model-independent observation of odderon exchange Ref,~\cite{Csorgo:2019ewn} in a significant manner: 
the exclusion of these first three D0 data points at $\sqrt{s} = 1.96$ TeV  minimally modifies our earlier result of the odderon significance given in Ref.~\cite{Csorgo:2019ewn}. We have repeated the analysis of Ref.~\cite{Csorgo:2019ewn} by leaving out the overlapping acceptance, by removing the
first 3 D0 points, but in an otherwise unchanged manner as compared to  Ref.~\cite{Csorgo:2019ewn}, and we have found that the statistical significance of odderon
exchange decreased only slightly,  from 6.26 to 6.10 $\sigma$, remaining safely well above the 5 $\sigma$ discovery threshold. This is due to the fact that most of the signal of odderon exchange arises from the kinematic range of the diffractive minimum and maximum. A more detailed investigation about this $-t$ range dependence 
of odderon exchange has been presented by T. Cs\"org\H{o} in his invited talk at the XXXVth IWHEP conference~\cite{Csorgo:2023Prot}.

The $8\to 1.96$ TeV and $1.96\to 8$ TeV projections correspond to keeping the measured $x$ values at $\sqrt{s} = $ 1.96 and 8 TeV, respectively, and determining by interpolation the $H(x,s)$ scaling functions at these $x$ values, but at the other energy, $\sqrt{s} = $ 8 and 1.96 TeV, respectively. This procedure is described in full details in  Ref.~\cite{Csorgo:2019ewn}.

\section{Quantification of significance of Odderon-exchange}

The left panel of Fig.~\ref{fig:H(x)-Odderon-3} compares the $H(x)$ scaling function of elastic $pp$ collisions at $\sqrt{s} = 8$ TeV to that of $p\bar p$ collisions at $\sqrt{s} = 1.96$ TeV. In this case, adopting the method of Ref.~\cite{Adare:2008cg}, the confidence level of the agreement of the $H(x)$ scaling functions is found to be lower then $ 0.015$\%, with a minimum of $\chi^2/{\rm NDF} = 46.4/17$. Hence, the difference between these scaling functions is statistically significant,
an at least 3.79 $\sigma$ effect, and it represents our result for the odderon observation from the comparison of the $\sqrt{s}= 8$ TeV $pp$
and  $\sqrt{s}= 1.96$ TeV $p\bar p$ datasets  in the $5 \le x\lessapprox 20$ acceptance domain. This seems to be a conservative and robust result as we find that this value can only increase for variations both in the procedure itself and in the $\chi^2$ definition.

%\section{Validity of H(x) scaling at 1.96 TeV}

%As a cross-check, we have tested the validity of the $H(x)$ scaling
%at ISR energies and found that all the differential 
%cross sections of elastic $pp$ scattering, measured at the ISR energy range of $\sqrt{s} = 23.5$ -- $62.5$ 
%GeV \cite{Amaldi:1979kd,Breakstone:1984te}, can approximately be scaled to the same universal curve~\cite{Csorgo:2019ewn}.
%We have also studied the $H(x)$ scaling 
%for elastic $p\bar p$ collisions in the energy range of $\sqrt{s} = 0.546$ -- $1.96$ TeV and 
%found that in this case, the scaling is limited to the diffractive cone, $x \le 10$ only, where 
%$H(x)$ $\approx$ $\exp(-x)$, but in $p\bar p$ collisions the $H(x)$ scaling is strongly 
%and qualitatively violated for $x > 10$ values. However, the valid $H(x)$ scaling in $pp$ scattering allowed us to scale down the higher energy TOTEM $pp$ data from $\sqrt{s} = 8$ TeV to $\sqrt{s} = 1.96$ TeV and directly compare it to the D0 $p\bar p$ data.

The quantification of the significance of odderon is based on a method developed by the PHENIX 
collaboration in Ref.~\cite{Adare:2008cg} using a specific $\chi^2$ definition that effectively 
diagonalizes the covariance matrix. In the case considered by the PHENIX Collaboration in Appendix A of Ref.~\cite{Adare:2008cg}, the experimental 
data are compared to a theoretical calculation. In our analysis, we adapt the PHENIX 
method~\cite{Adare:2008cg} for a comparison of one set of data directly to another set of data, without using any theory or fitting functions. 
Following the PHENIX method, we classify the experimental errors of a given data set into three different types: (i) type A, 
point-to-point fluctuating (uncorrelated) systematic and statistical errors, (ii) type B errors 
that are point-to-point dependent, but 100\% correlated systematic errors, and (iii) type C errors, 
that are point-to-point independent, but fully correlated systematic errors
to evaluate the significance of correlated data~\cite{Adare:2008cg}, when the full covariance matrix is not publicly available.
Since the $t$-dependent systematic errors in TOTEM measurements are almost 100 \% correlated, 
we classified them as type B errors, while the $t$-independent overall normalization errors are type C errors, 
and the statistical errors are type A errors.

The covariance matrix has been published together with the differential cross-section of elastic scattering at $\sqrt{s} = 13 $ TeV by TOTEM~\cite{Antchev:2018edk}, and on this dataset, we have checked with a Levy series expansion method~\cite{Csorgo:2018uyp},
that a fit with the full covariance matrix and another fit with our adopted PHENIX method gave the same minimum, the same central values for the fit parameters and the same errors of these parameters, within one standard deviation. This suggests that indeed the two methods are equivalent in our case too, and we utilized and adapted the PHENIX method for the comparision of two datasets, where the covariance matrix of at least one of these datasets was not publicly available.

At $\sqrt{s} = 2.76$ TeV, in Ref.~\cite{Antchev:2018rec}, the TOTEM Collaboration published the $pp$ differential cross section data with separated type A and type B errors. 
The source of the TOTEM $pp$ differential cross section data, measured at $\sqrt{s} = 7$ TeV, is Ref.~\cite{TOTEM:2013lle}. In addition, the values of $|t|$ were determined together with their errors of type A and B  
as given in Table 5 of Ref.~\cite{TOTEM:2013lle} and Table 3 of Ref.~\cite{TOTEM:2013lle}. The $t$-independent, type C errors cancel from the $H(x)$ scaling functions, as they multiply both the numerator and the denominator of $H(x)$. 

 The D0 collaboration did not publish type B errors for its differential cross-section data  at $\sqrt{s} = 1.96 $ TeV~\cite{Abazov:2012qb}. We have thus fixed the correlation coefficient of these D0 type B errors to zero. 
The input values of the nuclear slope parameters $B$ and the elastic cross sections $\sigma_{\rm el}$ are summarized in Table~\ref{table:B-sigma},  together with the appropriate references.
The sources of the TOTEM $pp$ differential cross section data, measured at $\sqrt{s} = 8$ TeV, are Refs.~\cite{Antchev:2015zza, TOTEM:2021imi} with errors of type A and B. The second reference specifies data without horizontal errors, only the bin widths are given that significantly overestimate those errors. Therefore, we have performed calculations with more realistic but likely still well overestimated horizontal errors of half bin widths, as well, resulting in significances much higher 
as compared to the significances presented in Table ~\ref{tab:odsum}.

We define the significance of the agreement between the data set $D_1$ and the projection $D_{21} = D_2 \to D_1$ of 
data set $D_2$ to $D_1$  in their overlapping acceptance,
with the following $\chi^2$ definition \cite{Csorgo:2019ewn}:
\begin{eqnarray*}
\chi^2_{2 \rightarrow 1} & = & \sum_{j=1}^{n_{21}}
    \frac{ (d_1^j +\epsilon_{b,1} e_{B,1}^j - 
    d_{21}^j - \epsilon_{b,21} e_{B,21}^j)^2 }
    {({\tilde e}_{A,1}^j)^2 + ({\tilde e}_{A,21}^j)^2} +
    \epsilon_{b,1}^2 + \epsilon_{b,21}^2 \,, 
    \\
\tilde{e}_{A,k}^j  & = &  e_{A,k}^j \frac{d_k^j + 
\epsilon_{b,k} e_{B,k}^j}{d_k^j} \,, \\
%\quad
e_{M,k}^j  & = & \sqrt{(\sigma_{M,k}^j)^2 + (d^{\prime,j}_k)^2 (\delta_{M,k}^j x)^2}  \,,
\end{eqnarray*}
where $n_{21}$ is the number of data points $d_{21}^j$ in $D_{21}$ indexed by $j$, 
the same as in $D_1$ but remaining in the overlapping acceptance of $D_{1,2}$ sets, 
$e_{M,k}^j$, $k=1,21$, are the type $M = A,B$ errors found in terms of 
the type M vertical errors on data point $j$, 
$\sigma_{M,k}^j$, added in quadrature with the corresponding type M vertical errors that were evaluated from 
the corresponding errors on the horizontal axis $x$ with the scaled variance method, $d^{\prime,j}_k \delta_{M,k}^j x$, 
where $d^{\prime,j}_k$ 
stands for the numerical derivative of the measured quantity in data set $D_k$ at the 
point $j$ in the common acceptance and $\delta_{M,k}^j x$ is the $j$-dependent type M horizontal error. 
The overall correlation coefficients of the type B errors $e_{B,k}^j$
of $D_k$ data sets are  
denoted by $\epsilon_{b,k}$. These coefficients are usually unknown, therefore we perform a scan of them to find the minimum of the $\chi^2$, whose corresponding values are indicated on the plots.
%,  respectively.
%%%%%%%%%%%%%%%%%%%%%%%%%%%%%%%%%%%%%
\begin{table*}[htb]
\begin{center}
\begin{tabular}{l l l}
$\sqrt{s}$ (GeV)  &\,\,\,\, $\sigma_{\rm el}$ (mb)  &\,\,\, $B$ (GeV$^{-2}$) \\
\hline
 1960 ($p\bar{p}$)  &\,\,\,\, 20.2  $\pm$ $1.7^{A} $ $\pm$ $14.4\%^{C}$ [*]  
 & \,\,\, 16.86 $\pm$ $0.1^{A}$  $\pm$ $0.2^{A}$  ~\cite{Abazov:2012qb} \\
 2760 ($pp$)          &\,\,\,\, 21.8  $\pm$ $1.4^{A} $  $\pm$ $6.0\%^{C}$  ~\cite{Nemes:2017gut,Antchev:2018rec}
                     & \,\,\, 17.1  $\pm$ $0.3^{A}$ ~\cite{Antchev:2018rec}     \\
 7000 ($pp$)          &\,\,\,\, 25.43 $\pm$ $0.03^{A}$ $\pm$ $0.1^{B}$ $\pm$ $0.31^{C}$ $\pm$ $1.02^{C}$  ~\cite{TOTEM:2013lle}  
 & \,\,\, 19.89 $\pm$ $0.03^{A}$ $\pm$ $0.27^{B}$ ~\cite{TOTEM:2013lle} \\
 8000 ($pp$)          &\,\,\,\, 27.1 $\pm$ $1.4^{A}$ ~\cite{Antchev:2013paa}
                      & \,\,\, 19.9 $\pm$ $0.3^{A}$ ~\cite{Antchev:2013paa} \\
\hline
\end{tabular}
\end{center}
\caption {Summary table of the elastic cross-sections $\sigma_{\rm el}$ and the nuclear slope parameters $B$, 
with references. We have indexed with superscripts $A,B,C$ the type A,B,C errors, respectively. The value and the type A error of the elastic cross-section $\sigma_{\rm el}$ at $\sqrt{s} = 1.96$ TeV [*] 
is obtained from a low $-t$ exponential fit to the data of Ref.~\cite{Abazov:2012qb}, while the type C error 
is from Ref.~\cite{Abazov:2012qb}. The statistical and systematic errors of $d\sigma/dt$ 
data at $\sqrt{s} = 1.96$ TeV were  added in quadrature in Ref.~\cite{Abazov:2012qb}, therefore 
it was done in case of the elastic slope $B$ as well, providing a combined type A error 
$\delta^{A} B = 0.224$ GeV$^{-2}$. At $\sqrt{s} = 2.76 $ TeV, Ref.~\cite{Antchev:2018rec} provides the total error on $B$, 
without decomposing it into type A and type B parts. Similarly, the error 
on the TOTEM value of the elastic cross section 
at $\sqrt{s} = 2.76$ TeV was not decomposed to type A and B errors in Ref.~\cite{Nemes:2017gut}, either. Hence, we treat 
these as errors of type A: this assumption yields a conservative estimate 
of the Odderon significance in our calculations.
}
\label{table:B-sigma}
\end{table*}
%%%%%%%%%%%%%%%%%%%%%%%%%%%%%%%%%%%%%%

In the right panel of Fig.~\ref{fig:H(x)-Odderon-3} we present visible and statistically significant deviation from unity in the ratio of the
scaling functions  of $pp$ and $p\bar p$ elastic scattering.
The ratio of the $H(x)$ scaling functions is shown for elastic $p\bar p$ 
collisions at $\sqrt{s} = 1.96$ TeV over that of $pp$ collisions at $\sqrt{s} = 8$ TeV. As a cross-check, we show 
the results of two different projection procedures: direct $1.96\to 8$ TeV and inverse $8\to 1.96$ TeV denoted by blue 
and red circle points, respectively. No significant variation with respect to the direction of projection has been found. 
In both ways, we observe an Odderon effect as a peak in the $5<x<10$ region, followed by a factor 
of two suppression or decrease from unity in a broad range of $10\lessapprox x=-tB\lessapprox 20$. The statistical significance 
of the observed difference between the $pp$ and $p\bar p$ scaling functions has been found to be at least 3.79 $\sigma$, which corresponds to an indication but not alone a statistically significant proof of Odderon exchange from the comparison of these two datasets only. It is important to emphasize that this result is obtained by utilizing the scaling properties of $pp$ scattering without any reference to modelling and without removing (or adding to) any of the published D0 or TOTEM data points, that is all the available data (in particular all the 17 D0 points) were utilized in this analysis.
This is in contrast to the D0-TOTEM publication on observation of Odderon exchange~\cite{D0:2020tig}, where the analysis was limited to only 8 of all the 17 available D0 points~\cite{Abazov:2012qb}.

\section{Summary of Odderon-exchange significances}

The significances of the model-independent Odderon-exchange observations in framework of the $H(x)$ scaling analysis are summarized in Table~\ref{tab:odsum} based on the new results discussed in this paper and as well as on the ones in our previous paper, Ref.~\cite{Csorgo:2019ewn}.  One can notice that no Odderon signal is observed when the $H(x, s_1 | p\bar p)$ scaling function calculated from the $\sqrt{s_1} = 1.96$ TeV $p\bar{p}$ D0 data is compared to the $H(x,s_2 | pp )$ scaling function calculated from the $\sqrt{s_2} = 2.76$ TeV $pp$ TOTEM data. However, this small minimal significance value is due to the smaller acceptance domain in $-t$ of the 2.76 TeV data set which reflects the importance of having $pp$ data points possibly in the whole D0 acceptance. The $p\bar{p}$ 1.96 TeV vs. $pp$ 8 TeV $H(x)$ function comparison results in an Odderon signal that alone does not reach the 5 $\sigma$ discovery level significance limit.
However, the comparison of the $H(x |s_1, p\bar{p})$ of 1.96 TeV and the $H(x |s_3, p p)$ data at $\sqrt{s_3} = 7 $ TeV $H(x)$ itself results in 
a statistically significant signal of Odderon exchange, that is well above the 5$\sigma$ discovery level significance limit~\cite{Csorgo:2021epjc,Csorgo:2019ewn,Csorgo:2020msw}.  

What happens, if we combine significances obtained from various pairwise comparisons?
The combined significances of Odderon exchange using the $H(x)$ scaling analysis are summarised in Table~\ref{tab:odsum-combo}. For a combination of the significances two different methods are used: 1) summation of the individual $\chi^2$ and NDF values; 2) Stouffer's method (used also by TOTEM in Ref.~\cite{D0:2020tig}). It is  important to stress that, independently of the method used for a combination, the combined Odderon observation significance is well above the 5 $\sigma$ discovery level when all possible $pp$-$p\bar p$ $H(x)$ scaling function comparisons are taken into account.
\begin{table*}[hbt]
    \centering
    \begin{tabular}{cccccc}
%    \hline
        $\sqrt{s}$ (TeV) & $\chi^2$ & NDF & CL & significance ($\sigma$)   \\ \hline %\hline
        1.96  vs. 2.76 & 3.85 & 11 & 9.74$\times 10^{-1}$ & 0.03  \\ %\hline
        1.96 vs. 7 & 80.1 & 17 & 3.681$\times$10$^{-10}$ & 6.26 \\ %\hline
        1.96 vs. 8 & 46.4 & 17 & 1.502$\times$10$^{-4}$  & 3.79  \\ %\hline
    \end{tabular}
    \caption{Summary of the Odderon signal in the $H(x)$ scaling analysis.}\label{tab:odsum}
\end{table*}

\begin{table*}[hbt]
    \centering
    \begin{tabular}{cccccc}
  %  \hline
        $\sqrt{s}$  (TeV) & $\chi^2$ & NDF & CL & \shortstack{ \\ $\chi^2$/NDF method} & \shortstack{ combined $\sigma$\\ Stouffer's method}  \\ \hline %\hline
        1.96 vs  2.76 \& 8 & 50.25 & 28 & 6.064$\times 10^{-3}$ & 2.74 & 2.70   \\ %\hline
        1.96 vs  2.76 \&  7 & 83.95 & 28 &1.698$\times 10^{-7}$ & 5.22 & 4.44  \\ %\hline
        1.96 vs  2.76 \&  7 \&  8 & 130.35 & 45 & 2.935$\times 10^{-10}$ & 6.30 & 5.81   \\ %\hline
        1.96 vs  7 \&  8 & 126.5 & 34 & 1.415$\times 10^{-12}$ & 7.08 & 7.10   \\ %\hline
    \end{tabular}
    \caption{Combined Odderon significances. The individual $\chi^2$, NDF and $\sigma$ values of each collision energy are taken from Table~\ref{tab:odsum}. }\label{tab:odsum-combo}
\end{table*}

\section{Conclusion}
With the new $\sqrt{s}$ = 8 TeV $pp$ data of the TOTEM collaboration we have further strengthened the observation of Odderon that we have already shown in our previous analysis with 7 TeV data.
%The $H(x)$ analysis of the new 8 TeV data of $pp$ elastic scattering confirmed the existence of the Odderon with at least 3.79 $\sigma$ significance in this particular case. 
%If we combine our new result, u
Using the $\chi^2$ method, % obtained with the help of the new TOTEM $pp$ data at $\sqrt{s} = 8$ TeV with our earlier 6.26 $\sigma$ statistical significance for odderon-exchange, 
% obtained with earlier TOTEM $pp$ data at $\sqrt{s} = 7$ TeV,   
the combined significance of Odderon exchange grows from 6.26 to 7.08 $\sigma$,  as shown in Table~\ref{tab:odsum-combo}, as obtained from  TOTEM data on elastic   $pp$  scattering at 7 and 8 TeV and all D0 data on elastic $p\bar p$ scattering at 1.96 TeV. As also indicated in Table~\ref{tab:odsum-combo}, Stouffer's method for the combination of significances yields similar results, resulting in a 
7.10 $\sigma$ combined statistical significance of odderon exchange using all the D0 elastic $p\bar p$ data at 
$\sqrt{s} = 1.96$ TeV, Ref.~\cite{Abazov:2012qb} and all the 
TOTEM data at  $\sqrt{s} = 7$ and $\sqrt{s} = 8$ TeV, Refs.~\cite{TOTEM:2013lle,TOTEM:2021imi}. 
%{\color{red} Also, we have concluded that the small -t ATLAS data do not play a significant role in the determination of the Odderon in this analysis.}
Let us note, that the domain of validity of this $H(x,s|pp)$ scaling includes these three energies, but so far this property has been published only in a model-dependent way~\cite{Csorgo:2021epjc,Szanyi:2022ezh}. The domain of validity of this $H(x|pp)$ scaling has been found to include $\sqrt{s} = 1.96$, $2.76$, $ 7$ and $ 8$ TeV in a model-independent way as well as presented at various conferences~\cite{Csorgo:2023Prot}, but these results go well beyond the scope of the present manuscript and will be written up in full details in a separate publication.

%The first discovery of the Odderon was published in three papers by three different methods. This model-independent analysis has provided the earliest evidentiary results, now expanded with additional 8 TeV ones. This method is unique in the sense that the other two ones are either entirely or partially model-dependent and they do not use all the data points in each data sets. The next scientific goals are determination of the properties of the Odderon. 

\textit{Acknowledgments.} 
We acknowledge inspiring and useful discussions with W.~Guryn, G.~Gustafson, 
L. Jenkovszky, V.A.~Khoze, E.~Levin, L.~L\"onnblad, M.~Strikman as well as with  members of the D0 and TOTEM collaborations.
R.P.~is supported in part by the Swedish Research Council grant No.~2016-05996, by the European Research Council (ERC) under the European Union's Horizon 2020 research and innovation programme (grant agreement No 668679).
T.Cs., T.N., A.S. and I.Sz. were partially supported by the NKIFH grants  K-133046 and K-147557 as well as by the EFOP 3.6.1-16-2016-00001 grant (Hungary). I. Sz. was supported as well as by the ÚNKP-21-3 and ÚNKP-22-3 New National Excellence Program of the Hungarian Ministry for Innovation and
Technology from the National Research, Development and Innovation Fund.
Our collaboration has also been partially supported by the COST Action CA15213 (THOR) and by the MATE KKP 2023 $\&$ 2024 grants.

%%%%%%%%%%%%%%%%%
\bibliographystyle{spphys}  
\bibliography{ISMD2023_proc}
%%%%%%%%%%%%%%%%%

\end{document}